\documentclass[
 amsmath,amssymb,
 aps,
]{revtex4-2}

\usepackage{url}
\usepackage[%
  colorlinks=true,
  urlcolor=blue,
  linkcolor=blue,
  citecolor=blue
]{hyperref}

\usepackage{placeins} 
\usepackage{xcolor}
\usepackage{xparse}

\usepackage{graphicx}
\usepackage{bm}

\begin{document}


\title{Double transition in kinetic exchange opinion models with activation dynamics}

\author{Marcelo A. Pires$^{1}$}
\thanks{piresma@cbpf.br}

\author{Nuno Crokidakis$^{2}$}
\thanks{nuno@mail.if.uff.br}

\affiliation{
$^{1}$Departamento de Física, Universidade Federal do Ceará, Fortaleza/CE, Brazil
\\
$^{2}$Instituto de F\'isica, Universidade Federal Fluminense, Niter\'oi/RJ, Brazil
}

\date{\today}

\begin{abstract}
In this work we study a model of opinion dynamics considering activation/deactivation of agents. In other words, individuals are not static and can become inactive and drop out from the discussion. A probability $w$ governs the deactivation dynamics, whereas social interactions are ruled by kinetic exchanges, considering competitive positive/negative interactions. Inactive agents can become active due to interactions with active agents. Our analytical and numerical results show the existence of two distinct nonequilibrium phase transitions, with the occurrence of three phases, namely ordered (ferromagnetic-like), disordered (paramagnetic-like) and absorbing phases. The absorbing phase represents a collective state where all agents are inactive, i.e., they do not participate on the dynamics, inducing a frozen state. We determine the critical value $w_c$ above which the system is in the absorbing phase independently of the other parameters. We also verify a distinct critical behavior for the transitions among different phases.

\end{abstract}

\maketitle

\section{\label{sec:intro}Introduction}

Opinion dynamics is one of the hottest topics of research in statistical physics of complex systems  \cite{2009castellanoFL, galam2008sociophysics,sirbu2017opinion,sznajd2021review,noorazar2020recent,conte2012manifesto}. One of the reasons for this interest is that even simple models can exhibit a complex collective behavior that emerges from the interactions among individuals. Usually, those models exhibit phase transitions and rich critical phenomena, which justifies the theoretical interest of physicists in the study of opinion dynamics.

In most opinion dynamics' models all agents are permanently active and have chances to interact with other agents \cite{vazquez2020role,perez2020opinion,crokidakis2019emergence,oestereich2020hysteresis}. However, many social environments especially online communities are not static in a way that agents can become inactive and withdraw from the discussion. In other words, social users do not concentrate on the discussion all the time, and they may lose interest and drop out of it \cite{xiong2011opinion,toscani}. However, these dormant agents can become active again following peers \cite{xiong2011dissipative}. Some works have verified that the final average opinion depends significantly on the external influence and internal actions. Small external activation drives the initially inactive system to total consensus quickly, but large external deactivation is required to freeze the active dynamics \cite{liu2013external}. This kind of activation dynamics leads to interesting results. The authors in \cite{xiong2011opinion} verified that if dissipation stays below a threshold value, the system evolves to a balance (paramagnetic) state where the average concentration of one opinion is equal to that of the other. On the other hand, it was verified that, in activation-like opinion dynamics that evolves on the top of complex networks, the topology of network also evolves in time \cite{xiong2011dissipative}. The authors verified that the model has power-law degree distribution, and clustering coefficients stay higher than results in Barabási-Albert networks. Another study revealed that under the impact of external circumstances, the population can evolve to distinct stationary states. On one hand, one opinion can finally be made dominant when the internal motivation is sufficiently large. On the other hand, without external activation, consensus is hardly reached in the system with interest decay \cite{xiong2011dissipative}.

On the other hand, we have the kinetic exchange opinion models (KEOM), that have been subject of study since the work of Lallouache-Chakrabarti-Chakraborti-Chakrabarti (LCCC) \cite{2010lallouacheCCC}. The LCCC model introduced a dynamic rule for opinion dynamics based on models of wealth exchange. The model considered continuous opinions in the range [-1.0,1.0], but discrete opinions were also considered in \cite{2011biswas}. Several extensions for discrete and continuous opinions were also studied, considering for example the impact of agents' convictions \cite{2012crokidakisA}, social temperature \cite{2017anteneodoC}, inflexibility \cite{crokidakis2014impact}, nonconformist behaviors \cite{2016vieiraAC}, dynamic individual influence \cite{xiong2014kinetic}, the presence of contrarian individuals \cite{joao2017}, influencing ability of individuals \cite{2011sen,2012biswasCS}, the relation between coarsening and consensus \cite{mukherjee2020long}, competition between noise and disorder \cite{crokidakis2016noise}, the analysis of noise-induced absorbing phase transitions \cite{vieira2016noise} and the presence of distinct interaction rules \cite{2011biswasCC}. The model was also considered in finite dimensional lattices \cite{2016mukherjeeC} (including applications to the 2016 presidential election in USA \cite{biswas2017critical}), in triangular, honeycomb, and Kagome lattices \cite{lima2017nonequilibrium}, in quasiperiodic lattices \cite{alves2020consensus}, in modular networks \cite{2019oestereichPC} and in other complex networks \cite{liu2018kinetic}.

Take into account those two subjects, namely activation dynamics in opinion formation and kinetic exchange opinion models, we propose the inclusion of activation dynamics considering the KEOM social interaction rules. Coevolution spreading were considered in other works \cite{wang2019coevolution,pires2017dynamics,2018piresOC,PhysRevE.104.034302,ferrari2021coupling,wang2015coupled,Feng2017,doi:10.1142/S0129183109013728,DING20101745,Liu_2019}. Here we consider that the activation dynamics follows a contact-like process \cite{marro2005nonequilibrium}, and the social interactions are ruled by the KEOM discussed in \cite{2012biswasCS}.

This work is organized as follows. In section 2 we discuss our coevolution dynamics, and present the microscopic rules that govern our model. The analytical and numerical results are discussed in section 3. Finally, the conclusion and final remarks are presented in section 4.


\section{Model}

We consider a fully-connected population with $N$ agents.  The agents can be classified as follows: 
\begin{itemize}
\item[$(I)$] Opinion: $o_i =  +1$ if an agent $i$ supports opinion $+1$, 
$o_i =  -1$ if $i$ supports opinion $-1$, 
$o_i =  0$ if $i$ is undecided/neutral.

\item[$(II)$] Activation status: $s_i =  1$ if an agent $i$ is active or $s_i =  0$ if $i$ is inactive. 
\end{itemize}

Considering the social interactions, as explained below, we will follow the Biswas-Chatterjee-Sen (BCS) model \cite{2012biswasCS}. In the BCS model, two agents $i$ and $j$ are randomly chosen, and they interact through the following rule:
\begin{equation} \label{eq:evol}
o_i(t+1) = o_i(t) + \mu_{ij}\,o_j(t)    ~.
\end{equation}
\noindent
This expression shows how the opinion of a given agent $i$ in a given time step $t+1$ is updated due to a interaction with another agent $j$. The first term in the right side of the equation indicates the tendency of agent $i$ to keep his/her current opinion in the time step $t$, but the opinion can be influenced by another agent $j$. The coupling $\mu_{ij}$ represents the strength of the pairwise interaction. Pairwise interaction strengths are annealed random variables distributed according to the binary probability density function (PDF)
\begin{equation}\label{mu}
F(\mu_{ij})=p\,\delta(\mu_{ij}+1) + (1-p)\,\delta(\mu_{ij}-1)  ~.  
\end{equation}
In other words, the agents can exchange opinions with positive ($+1$) or negative ($-1$) influences, and $p$ quantifies the mean fraction of negative ones. Notice that, in Eq. \eqref{eq:evol}, if the value of the opinion exceeds (falls below) the value $1\,(-1)$, then it adopts the extreme value $1 \,(-1)$ \cite{2010lallouacheCCC}. The model defined by Eqs. \eqref{eq:evol} and \eqref{mu} presents only one parameter, $p$, and it undergoes an order-disorder phase transition at a critical value $p_c=1/4$, i.e., the competitive positive/negative interactions are responsible for such ferromagnetic-paramagnetic transition \cite{2012biswasCS}.

In this work we propose the inclusion of the above-mentioned activation state  of agents (see above rule II) in the BCS model. In such a case, at each time step the following rules govern the dynamics of our model:
\begin{itemize}
\item Select two agents $i$ and $j$;

\item If $i$ is active: 

 \begin{itemize}
 \item  with probability $w$ apply the deactivation process: $ s_i =  1 \rightarrow s_i =  0$. \textbf{(Rule 1)}. 	
 
 \item with probability $1-w$, the agents $i$ and $j$ follow the BCS model: 
 $o_i(t+1) = o_i(t) + \mu_{ij} o_j(t) $, where  $\mu_{ij}$ follows the PDF of Eq. \eqref{mu} \textbf{(Rule 2)}.
  \end{itemize}

\item If $i$ is inactive: 
 \begin{itemize}
 \item  if  $i$ and $j$ have the same opinion ($o_i = o_j$) and $j$ is active: apply  the activation process $ s_i =  0 \rightarrow s_i =  1$  with probability $(1-p)g$. 
  \textbf{(Rule 3)}. 
  \end{itemize}
  
  \end{itemize}

As discussed in the Introduction, during a public debate there are people that that are not active/open for discussion. But they can become active due to peers influence. As discussed in \cite{xiong2011opinion,liu2013external}, active agents may gradually lose their interests in the discussion and drop out of it, which is related to our above \textbf{Rule 1}. In addition, our above \textbf{Rule 2} is based on the fact that only active agents participate in the opinion dynamics. Finally, as also discussed in \cite{xiong2011opinion,liu2013external}, active agents can motivate their inactive like-minded peers. If an inert agent holds the same opinion of an active neighbor and they have a positive interaction (with takes place with probability  $1-p$) then, the inert agent is driven to be active with probability $(1-p)g$. In other words, in order to have an activation with probability $g$, we considered as a necessary condition that the randomly chosen agents $i$ and $j$ have to interact positively, which occurs with probability $1-p$, leading to the final probability $(1-p)g$. These last sentences justify our above \textbf{Rule 3}.


\section{Results}

In a given time step $t$, let $x_{+1}(t) $, $x_{0}(t) $ and $x_{-1}(t) $ be the proportion of \textbf{inactive} agents with opinions $+1$, $0$ and $-1$, respectively. Also, let $f_{+1}(t) $, $f_{0}(t) $ and $f_{-1}(t) $ be the proportion of \textbf{active} agents with opinions $+1$, $0$ and $-1$, respectively. Following the analytical procedure of refs. \cite{2011biswas,2012biswasCS,2012crokidakisA} and the rules defined in the previous section (\textbf{Rules 1, 2} and \textbf{3}), we can write the master equations for the evolution of the fractions of inactive agents as follows:

\begin{equation}
\text{ (Inactive with opinion +1)} \quad\quad \frac{d x_{+1} }{dt}  
= 
w f_{+1} - (1-p)g  x_{+1} f_{+1} 
\end{equation}
\begin{equation}
\text{ (Inactive with opinion 0)} \quad\quad 
\frac{d x_{0} }{dt}  
= 
w f_{0} - (1-p)g  x_{0} f_{0} 
\end{equation}
\begin{equation}
\text{ (Inactive with opinion -1)} \quad\quad 
\frac{d x_{-1} }{dt}  
= 
w f_{-1} - (1-p)g  x_{-1} f_{-1} 
\end{equation}

In addition, we have the following master equations for the evolution of the fractions of active agents with opinions $o=+1$ and $o=-1$:

\begin{equation}\label{Eq.dfplus.dt}
\frac{d f_{+1} }{dt}  
= 
-w f_{+1} + (1-p)g  x_{+1} f_{+1}
+(1-w)
\left[
 -p f_{+1}^2 - (1-p) f_{+1}f_{-1} 
 +(1-p) f_{+1}f_{0}  + p f_{-1}f_{0} 
\right] 
\end{equation}
\begin{equation}\label{Eq.dfminus.dt}
\frac{d f_{-1} }{dt}  
= 
w f_{-1} - (1-p)g  x_{-1} f_{-1}
+(1-w)
\left[
 -p f_{-1}^2 - (1-p) f_{-1}f_{+1} 
 +(1-p) f_{-1}f_{0}  + p f_{+1}f_{0} 
\right] 
\end{equation}

Additionally, we have the normalization condition:
\begin{equation} \label{normaliz}
x_{+1}+x_{0}+x_{-1}
+
f_{+1}+f_{0}+f_{-1}
=1
\end{equation}

In the steady state, we have for Eqs. (3), (4) and (5): 
\begin{equation} \label{eq_x1}
0=\frac{d x_{+1} }{dt}  
= 
w f_{+1} - (1-p)g  x_{+1} f_{+1}
\Rightarrow 
f_{+1} = 0 \text{ or }
x_{+1} = \frac{w}{(1-p)g}
\end{equation}
\begin{equation}\label{eq_xo}
0=\frac{d x_{0} }{dt}  
= 
w f_{0} - (1-p)g  x_{0} f_{0}
\Rightarrow 
f_{0} = 0 \text{ or }
x_{0} = \frac{w}{(1-p)g}
\end{equation}
\begin{equation}\label{eq_xm1}
0=\frac{d x_{-1} }{dt}  
= 
w f_{-1} - (1-p)g  x_{-1} f_{-1}
\Rightarrow
f_{-1} = 0 \text{ or }
x_{-1} = \frac{w}{(1-p)g}
\end{equation}

Thus the fraction of active agents, $\rho=f_{+1}+f_{0}+f_{-1}$, in the steady state is obtained from the normalization condition Eq. \eqref{normaliz} written in the form $f_{+1}+f_{0}+f_{-1}=1-(x_{+1}+x_{0}+x_{-1})$. Considering the results \eqref{eq_x1}, \eqref{eq_xo} and \eqref{eq_xm1} for $x_{+1}, x_{0}$ and $x_{-1}$, respectively, we have
\begin{equation}
\rho=1-\frac{3w}{(1-p)g}
\label{rho}
\end{equation}
The importante of such fraction $\rho$ will be discussed in the following.

We are interested in the critical behavior of the model. Thus, let us discuss about the order parameter, $m$. It is sensitive to the unbalance between extreme opinions $+1$ and $-1$. Notice that $m$ plays the role of the ``magnetization per spin'' in magnetic systems \cite{2012crokidakisA}. Since the order parameter can be defined as  $m=| f_{+1} -f_{-1}|$, we have
\begin{equation}
0=\frac{d m }{dt}  \Rightarrow
\frac{d f_{+1} }{dt}  
=
\frac{d f_{-1} }{dt}  
\Rightarrow
(f_{+1}-f_{-1})
\left[
p(f_{+1}+f_{-1})
+
(1-2p)f_{0}
\right] = 0 
\end{equation}
Then
\begin{equation} \label{paramag}
f_{+1}=f_{-1} \Rightarrow
 m = 0
\end{equation}
which represents a disordered state solution, or 
\begin{equation}
f_{+1}+f_{-1} = \frac{1-2p}{p} f_{0} 
\label{Eq.soma.f.2}
\end{equation}

Using that $f_{+1}+f_{-1}=\rho-f_{0}$ we get
\begin{equation}
f_{0}  =  \frac{p}{1-p}\rho 
\label{Eq.fo_rho}
\end{equation}

Inserting  Eq. \eqref{Eq.fo_rho} into  Eq. \eqref{Eq.dfplus.dt} and using Eq. \eqref{eq_x1} we obtain 
 
\begin{equation}
f_{+1}^2 
-\rho\frac{1-2p}{1-p} f_{+1}
+\rho^2 \frac{p^2}{(1-p)^2} = 0 
 \label{eq.quadratica_fplus}
\end{equation}
Thus,
\begin{equation}
f_{+1} = \frac{\rho}{2(1-p)} \left( 1-2p \pm \sqrt{1-4p} \right)
 \label{eq.quadratica_fplus_2}
\end{equation}

Using that $f_{-1}= \rho-f_{0}-f_{+1}$ and Eqs. \eqref{rho} and \eqref{Eq.fo_rho} we arrive at

\begin{equation}\label{order_par}
m = | f_{+1} -f_{-1}| =\left(1-\frac{3w}{(1-p)g} \right)  \frac{\sqrt{ 1-4p } }{1-p}
\end{equation}

In the language of critical phenomena \cite{marro2005nonequilibrium}, one can rewrite Eq. \eqref{order_par} as
\begin{align} 
m \sim
\left( p_{ca}-p \right)^{\beta_{cp}}
\left( p_{ci}-p \right)^{\beta_{ising}}
\label{Eq.m.critico}
\end{align}
where $\beta_{cp}=1$ and $\beta_{ising}=1/2$ and 
\begin{equation} 
p_{ci} = \frac{1}{4}
 \quad\quad 
p_{ca} =  1-\frac{3w}{g}
\label{Eq.limiar}
\end{equation}

Eq. \eqref{Eq.m.critico} predicts the occurrence of two distinct nonequilibrium phase transitions. The critical point $p_{ci}$ is the same for the BCS model, i.e., $p_{ci}=1/4$. For this first critical point, the critical behavior suggests an Ising-like exponent $\beta=\beta_{ising}=1/2$, as in the BCS model \cite{2012biswasCS}. On the other hand, the second critical point is related to the activation dynamics, and it depends on the parameters $g$ and $w$, $p_{ca}= 1-\frac{3w}{g}$. For this second transition,  the critical behavior suggests a contact process-like exponent $\beta=\beta_{cp}=1$. Another form to see this second transition is also considering as another order parameter the fraction of active agents, Eq. \eqref{rho}, written in the form $\rho \sim \left( p_{ca}-p \right)^{\beta_{cp}}$. We will discuss those points in more details in the following.

\begin{figure*}[t]
\centering
\includegraphics[width=0.45\textwidth]{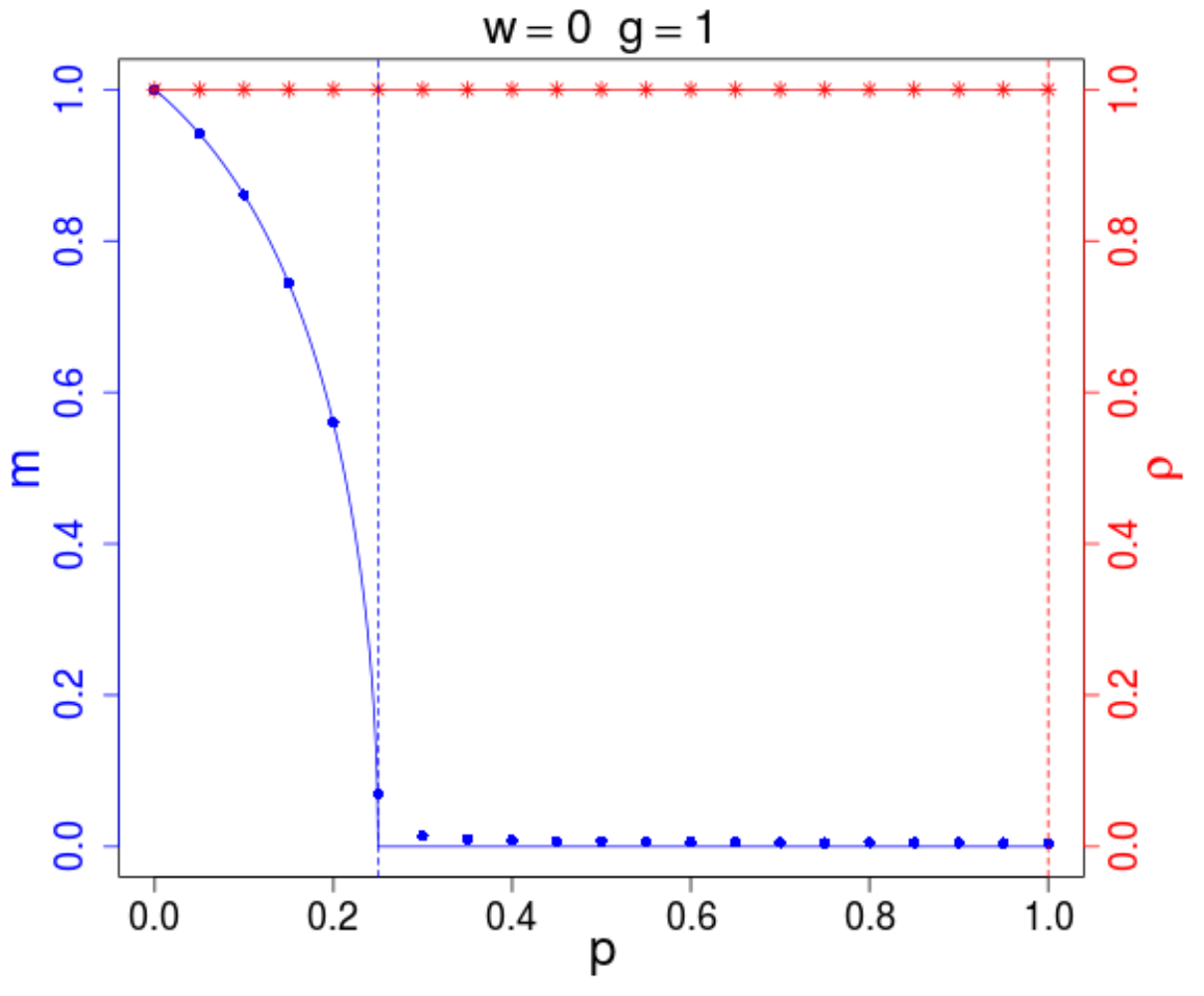}
\includegraphics[width=0.45\textwidth]{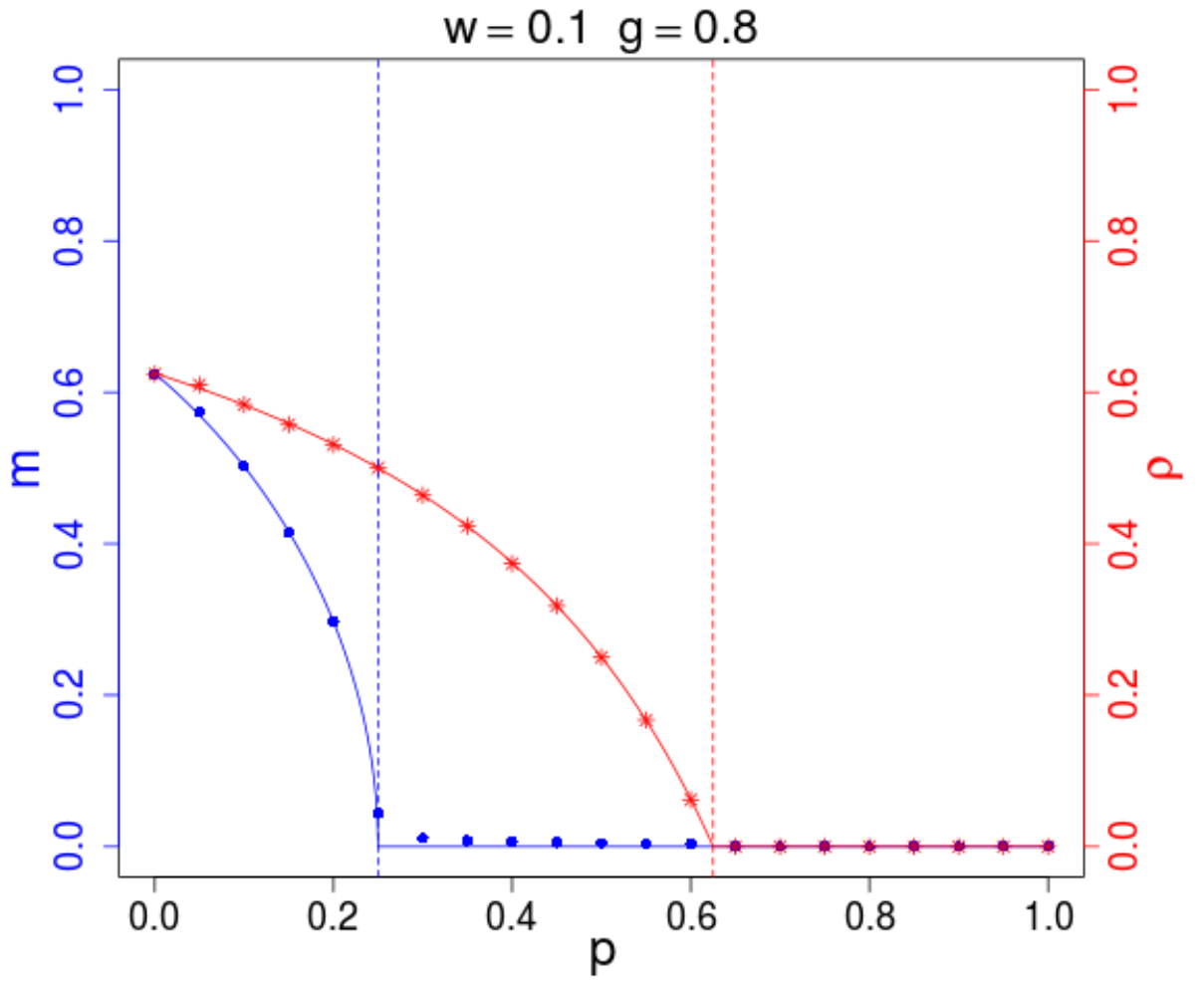}
\caption{Stationary order parameters $m$ (collective opinion)  and $\rho$ (fraction of active agents). (a) $w=0$ (the limiting case of the BCS model \cite{2012biswasCS}). (b) $w = 0.1$, $g=0.8$. Lines from Eqs. \eqref{order_par} and \eqref{rho}, and the vertical dashed lines are $p_{ci}=1/4$ and $p_{ca} =  1-\frac{3w}{g}$. Symbols come from Monte Carlo simulations following an agent-based protocol described in the section 2. The population size is $N=10^4$ agents, and the initial fraction of active agents is $\rho(0)=1$.}
\label{fig:mcs}
\end{figure*}

Based on the rules defined in section 2, we performed Monte Carlo simulations of the model, in order to confirm our analytical predictions. From the simulations, we can obtain $m$ through the definition
\begin{equation} \label{m_mc}
m = \left\langle \frac{1}{N}\left|\sum_{i=1}^{N} o_{i}\right|\right\rangle ~, 
\end{equation}
\noindent
where $\langle\, ...\, \rangle$ denotes average over  disorder or configurations, computed at the steady states. Usually in opinion dynamics models, $m$ is called collective opinion. In Fig. \ref{fig:mcs} we exhibit numerical results for $w=0.0, g=1.0$ (left side) and $w=0.1, g=0.8$ (right side). The lines for the order parameter $m$ were obtained from our analytical result, Eq. \eqref{order_par}, whereas the symbols were obtained from the simulations. In addition to the order parameter $m$, obtained numerically from Eq. \eqref{m_mc} (black symbols in Fig. \ref{fig:mcs}), we also measured in the simulations the fraction of active agents (red symbols in Fig. \ref{fig:mcs}). This last result is compared with the analytical result obtained from Eq. \eqref{rho}.

\begin{figure*}[t]
\centering
\includegraphics[width=0.48\textwidth]{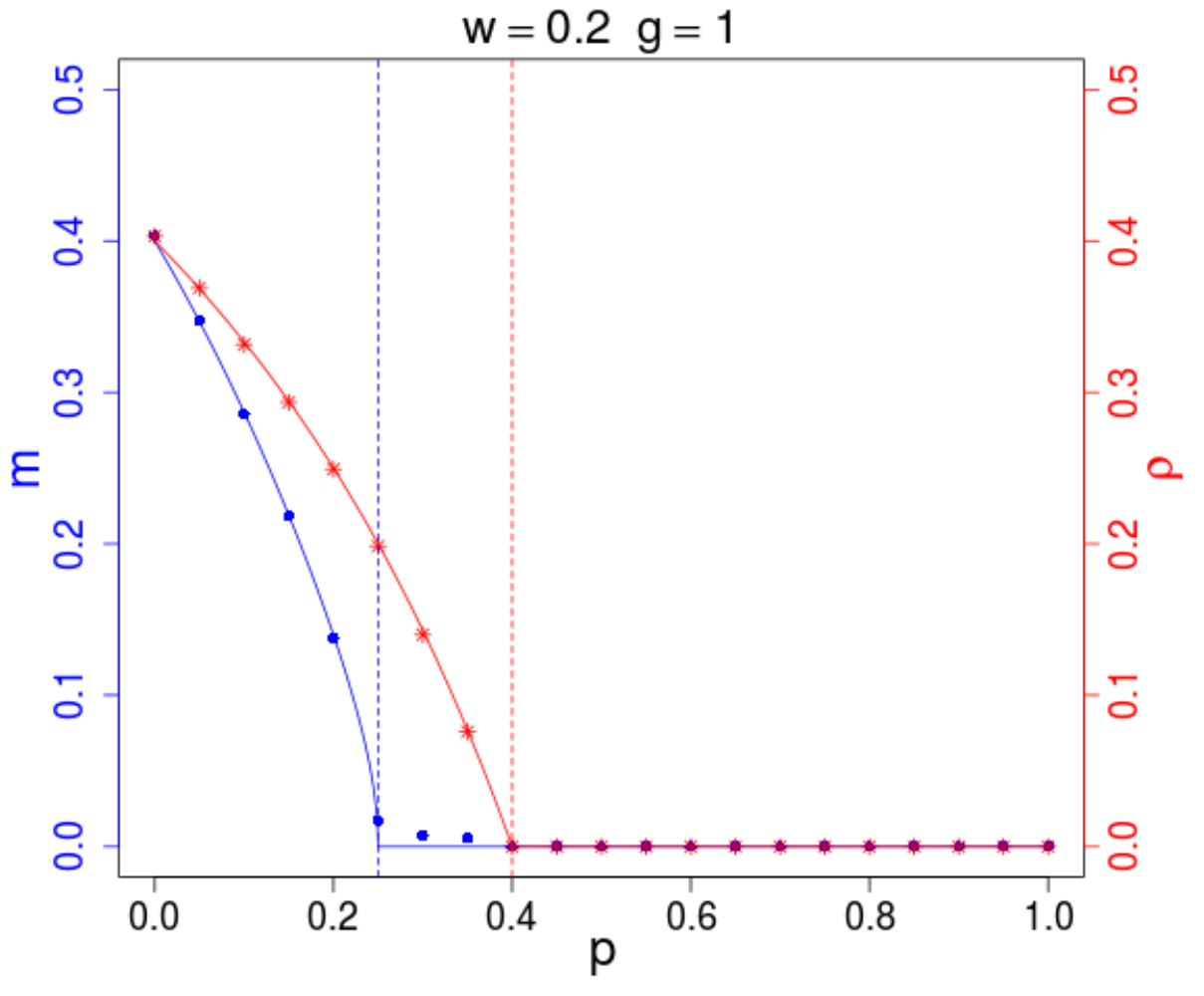}
\includegraphics[width=0.48\textwidth]{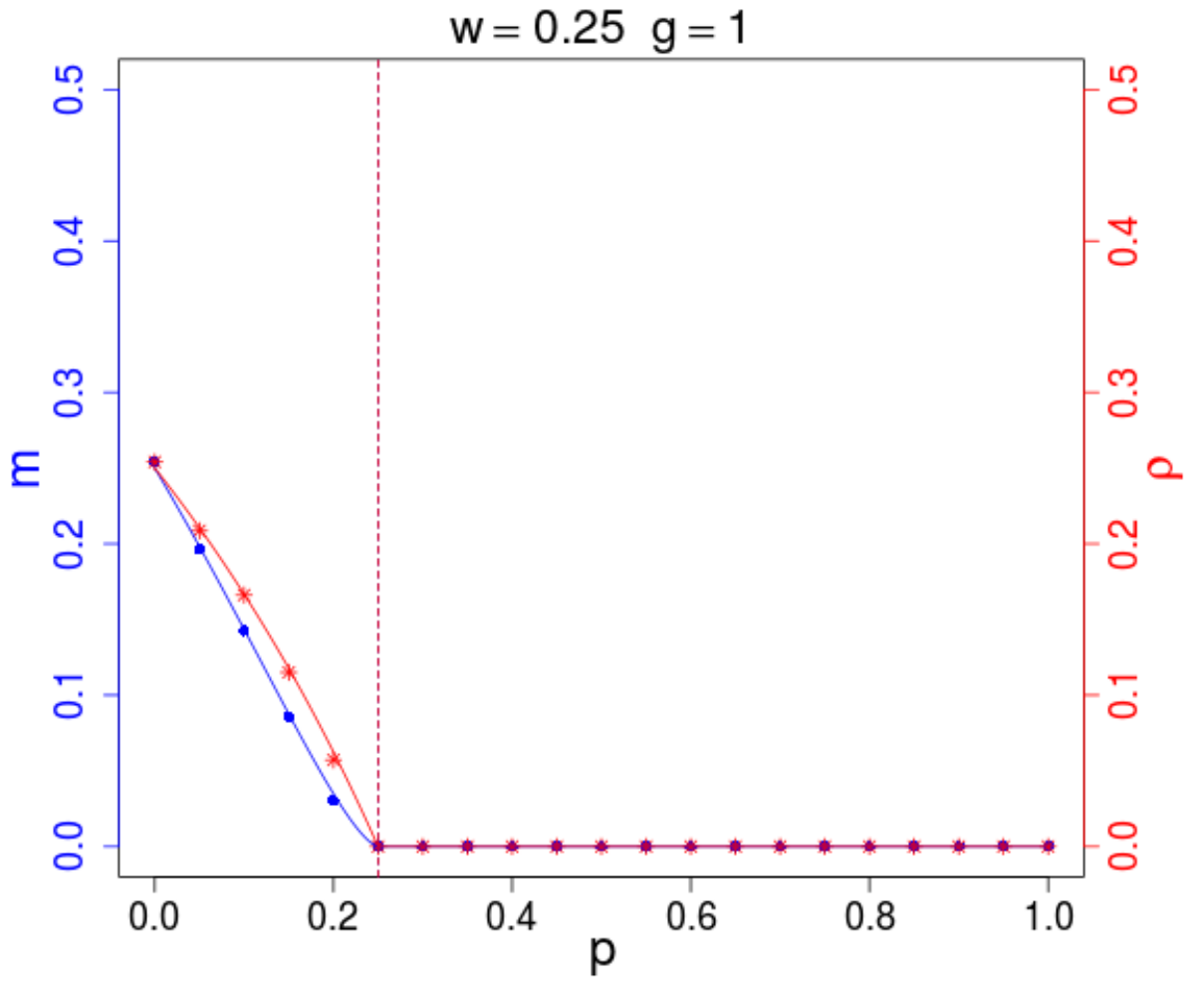}
\includegraphics[width=0.48\textwidth]{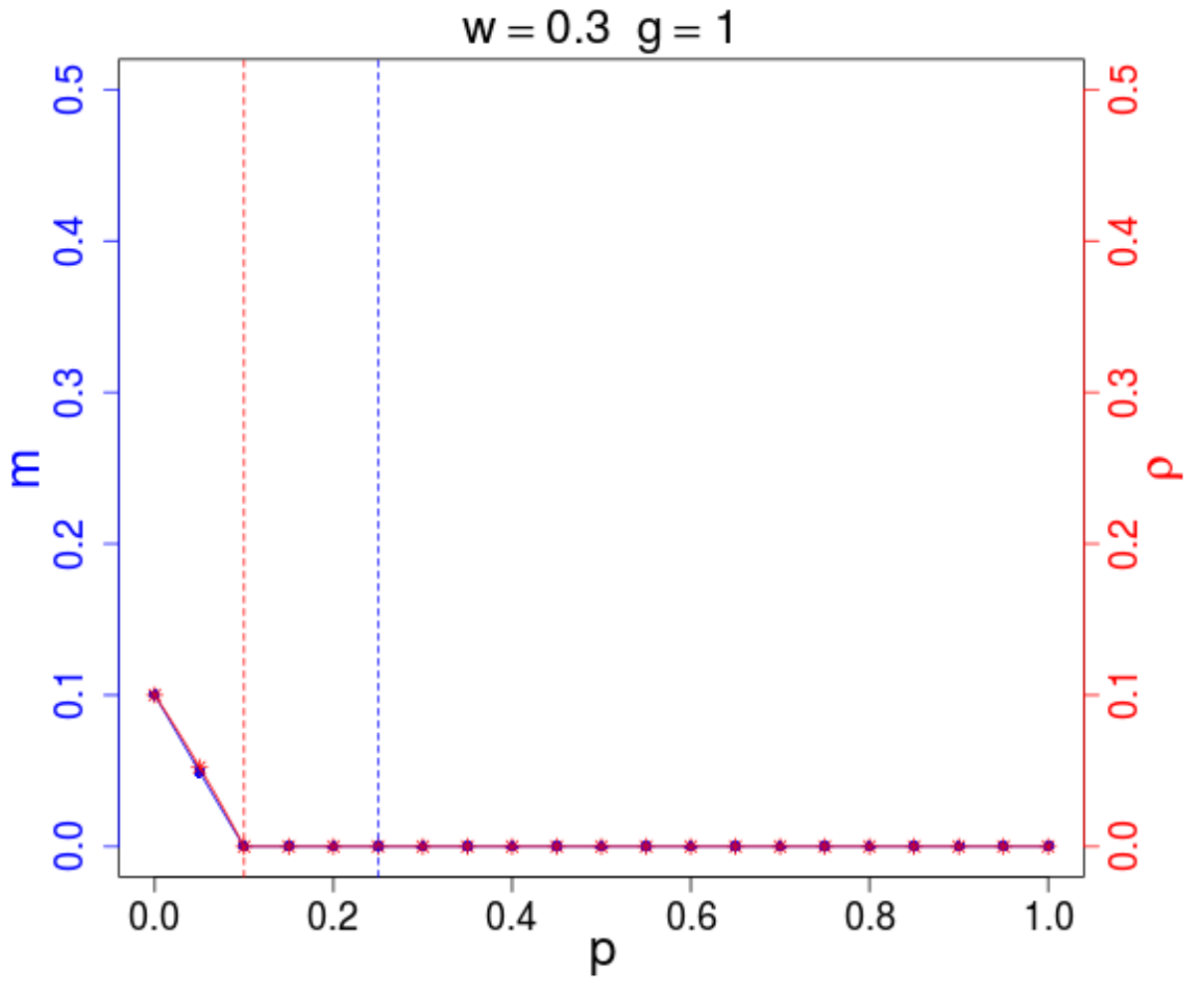}
\caption{Stationary order parameters $m$ (collective opinion)  and $\rho$ (fraction of active agents), for $w=\{0.20,0.25,0.30\}$. Solid lines are the analytical results (Eqs. \eqref{order_par} and \eqref{rho}). Symbols come from Monte Carlo simulations. The vertical dashed lines are $p_{ci}=1/4$ and $p_{ca} =  1-\frac{3w}{g}$. The population size is $N=10^4$ agents, and the initial fraction of active agents is $\rho(0)=1$.}
\label{fig:mcs2}
\end{figure*}

First of all, we exhibit in Fig. \ref{fig:mcs} (left side) the behavior of $m$ and $\rho$ for the special case $w=0$, for which we have no deactivation and thus we have to recover the results of Ref. \cite{2012biswasCS}. In such a case, for the order parameter $m$ we observed the nonequilibrium phase transition at $p_c=1/4$ studied in \cite{2012biswasCS}, and the fraction of active individuals is $\rho=1$ for all values of $p$, as expected since we have no deactivation. On the other hand, in Fig. \ref{fig:mcs} (right side) we exhibit the case $w=0.1$ and $g=0.8$. In such a case, we observed the abovementioned two distinct transitions: one for the order parameter $m$ and another for the fraction of active agents $\rho$. The critical points are obtained from Eq. \eqref{Eq.limiar}, i.e., the first critical point is not modified by the deactivation process and we have $p_{ci} = \frac{1}{4}$. However, due to the dynamics of activation/deactivation, the maximum value of the order parameters is less than $1$, as previous indicated by our analytical results, Eq. \eqref{order_par}. This is due to the presence of deactivated agents, i.e., we have now $x_{+1}>0, x_{-1}>0$ and $x_{0}>0$, which leads to smaller values of the active fractions $f_{+1}$ and $f_{-1}$ and consequently we have lower values of the collective ordering measure $m$, even for $p=0$. On the other hand, the process of deactivation induces a second critical point that depends on $w$ and $g$, given by $p_{ca} =  1-\frac{3w}{g}$. For the parameters we considered for Fig. \ref{fig:mcs}, we have $p_{ca} = 0.625$. The two nonequilibrium critical points are indicated by vertical lines in Fig. \ref{fig:mcs}.

Notice that the two transitions are of distinct nature. The analytical results of Eqs. \eqref{rho}, \eqref{Eq.m.critico} and \eqref{Eq.limiar} suggest two distinct critical exponents $\beta$. One of them is related to the behavior of the magnetization near the critical point $p_{ci}=1/4$, i.e., we have  $\beta=\beta_{ising}=1/2$. This transition is a usual ferromagnetic-paramagnetic transition. Indeed, we obtained a paramagnetic (disordered state) solution, Eq. \eqref{paramag}. In addition, we can observe Fig. \ref{fig:mcs} the usual finite-size effects for numerical results regarding ferromagnetic-paramagnetic phase transitions for values of $m$ near $p=p_{ci}$. On the other hand, for the second phase transition observed for $\rho$ at $p=p_{ca}=1-3w/g$, the analytical results suggest an active-absorbing phase transition, since the critical exponent is $\beta=\beta_{cp}=1$ \cite{marro2005nonequilibrium}. Indeed, we also obtained an analytical result $f_{+1}=f_{0}=f_{-1}=0$, see Eqs. \eqref{eq_x1}, \eqref{eq_xo} and \eqref{eq_xm1}, i.e., it is the absorbing state where $\rho=0$. In this case, the fraction of active agents needs to be zero even in the computer simulations, which in fact we observed (see the red points in Fig. \ref{fig:mcs}, right side).

\begin{figure*}[h]
\centering
\includegraphics[width=0.49\textwidth]{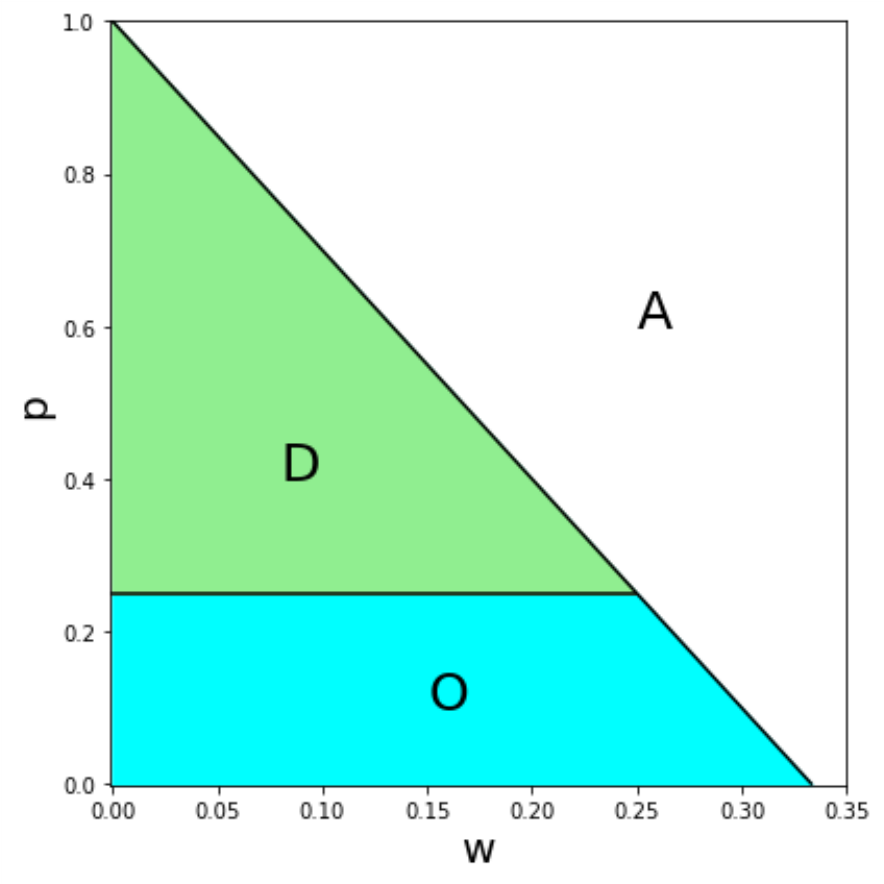}
\caption{Phase diagram of the model in the plane $p \times w$. Lines were obtained from Eq. \eqref{Eq.limiar} with the activation rate $g=1$.  Acronyms: O=Ordered phase, D=Disordered phase, A=Absorbing phase. We see the presence of multiple transitions: $O\rightarrow D\rightarrow A$ as well as $O\rightarrow A$.  For $w>w_c\approx 0.33$ there is no transition, as discussed in the text. Even with a small interest decay rate $w$ the  dynamics can enter in the absorbing state if the disagreement rate $p$ is high enough.} 
\label{fig:diagram-p-w}
\end{figure*}

In Fig. \ref{fig:mcs2} it is shown additional results for $m$ and $\rho$ for $g=1$ and typical values of the deactivation probability $w$. In Fig. \ref{fig:mcs2} (upper figure, left side) we exhibit the curves for $m$ and $\rho$ for $w=0.2$. As previous discussed, the maximum value of the magnetization decreases for increasing values of $w$, as predicted by Eq. \eqref{order_par}. In addition, the second critical point $p_{ca}$ decreases its value and becomes near the first critical point $p_{ci}$. One can also see from Eq. \eqref{Eq.limiar} that both critical points, $p_{ci}$ and $p_{ca}$ coalesce for a given value of $w$. Taking $p_{ci}=p_{ca}$, we obtain that such value of $w$ is given by $w = \frac{1}{4}\,g$. For the special case $g=1$, this equation gives us $w=0.25$. For this value we have $p_{ci}=p_{ca}=1/4$, which is exhibited in Fig. \ref{fig:mcs2} (upper figure, right side). The other result, $w=0.3$ suggests that there is a critical value of the deactivation dynamics above which there is no phase transition, and the system is in an absorbing state for all values of $p$. This critical value $w_{c}$ can be obtained from Eq. \eqref{Eq.limiar}, considering $p_{ca}(w_c)=0$. In this case, we obtain
\begin{equation} \label{wc2}
w_c=\frac{1}{3}\,g      
\end{equation}
For the special case $g=1$, Eq. \eqref{wc2} gives us $w\approx 0.33$. In Fig. \ref{fig:mcs2} (lower figure) we exhibit results for a value near such critical value $w_c$, namely $w=0.3$. We yet observe a transition, but the values of the order parameters $m$ and $\rho$ are quite small, indicating the proximity of the limiting case of occurrence of phase transitions, in agreement with Eq. \eqref{wc2}.

To summarize the results, we exhibit in Fig. \ref{fig:diagram-p-w} the phase diagram of the steady states of the model in the plane $p$ versus $w$, for $g=1$. Notice the critical value $w_c\approx 0.33$, above which the system is in the absorbing state for all values of $p$. It is also shown the Ordered (\textbf{O}) and Disordered (\textbf{D}) phases, and the transition between such phases is observed for the constant value $p_{ci}=1/4$. The other boundary is obtained by the second critical point $p_{ca}=1-3w/g$.


\section{Final Remarks}

During a public debate, individuals may abandon the discussion while others may become interested in an ongoing discussion by the influence of peers.  To model such a situation we consider an activation dynamics coupled to opinion dynamics. The activation/deactivation dynamics is ruled by a contact-like process, whereas the social interactions are governed by kinetic exchanges. The results show that such coupled dynamics undergoes multiple transitions, namely: 
(a) from an ordered state to a disordered state; 
(b) from a disordered state to an absorbing state;
(c) from an ordered state straight  to an absorbing state. Our mean-field  results suggest that the transition (a) takes place in the Ising universality class, whereas (b-c) occurs  in the universality class of the contact process. 
The absorbing state means that all the individuals stop participating in the discussion, so the debate fade out. Our results point out that even with a small interest decay rate $w$
 the  dynamics can be trapped in the absorbing state if the disagreement rate $p$ is too high. On the other hand, for sufficient high values of the deactivation rate $w$ the system is always in the absorbing phase, independently of the values of the other parameters. In future works it would be interesting to consider a networked extension of the model studied here as well as additional social features such as plurality and polarization\cite{oestereich2020hysteresis}.

\section*{Acknowledgments}

The  authors  acknowledge  financial  support  from  the  Brazilian  funding  agencies Fundação Cearense de Apoio ao Desenvolvimento Científico e Tecnológico (FUNCAP), Conselho Nacional de Desenvolvimento Cient\'ifico e Tecnol\'ogico (CNPq, grant number 310893/2020-8) and Funda\c{c}\~ao Carlos Chagas Filho de Amparo \`a Pesquisa do Estado do Rio de Janeiro (FAPERJ, grant number 203.217/2017).


\end{document}